# Testing earthquake predictions

Brad Luen[1] and Philip B. Stark[1]

*University of California, Berkeley*

**Abstract:** Statistical tests of earthquake predictions require a null hypothesis to model occasional chance successes. To define and quantify 'chance success' is knotty. Some null hypotheses ascribe chance to the Earth: Seismicity is modeled as random. The null distribution of the number of successful predictions – or any other test statistic – is taken to be its distribution when the fixed set of predictions is applied to random seismicity. Such tests tacitly assume that the predictions do not depend on the observed seismicity. Conditioning on the predictions in this way sets a low hurdle for statistical significance. Consider this scheme: When an earthquake of magnitude 5.5 or greater occurs anywhere in the world, predict that an earthquake at least as large will occur within 21 days and within an epicentral distance of 50 km. We apply this rule to the Harvard centroid-moment-tensor (CMT) catalog for 2000–2004 to generate a set of predictions. The null hypothesis is that earthquake times are exchangeable conditional on their magnitudes and locations and on the predictions–a common "nonparametric" assumption in the literature. We generate random seismicity by permuting the times of events in the CMT catalog. We consider an event successfully predicted only if (i) it is predicted and (ii) there is no larger event within 50 km in the previous 21 days. The $P$-value for the observed success rate is $< 0.001$: The method successfully predicts about 5% of earthquakes, far better than 'chance,' because the predictor exploits the clustering of earthquakes – occasional foreshocks – which the null hypothesis lacks. Rather than condition on the predictions and use a stochastic model for seismicity, it is preferable to treat the observed seismicity as fixed, and to compare the success rate of the predictions to the success rate of simple-minded predictions like those just described. If the proffered predictions do no better than a simple scheme, they have little value.

## 1. Introduction

Earthquake prediction has roots in antiquity [46]. Predictions have been based on a variety of seismic and non-seismic phenomena, including animal behavior [7, 18, 25, 26, 46]; water level, temperature and composition in wells and springs [2, 36]; electric and magnetic fields and radio waves on the ground and in the air [3, 50]; electrical resistivity of the ground and the air [3, 30, 53]; cloud formations or other atmospheric phenomena [41, 46]; infrared radiation [3]; pattern recognition [4, 24]; temporal clustering [11, 15]; and variations in the rate or pattern of seismicity [16].

There are large research efforts directed towards predicting earthquakes, such as the Collaboratory for the Study of Earthquake Predictability.[2] "Even a stopped clock is right twice a day," and almost any method for predicting earthquakes will succeed occasionally — whether the method has merit or not. Indeed, prominent

[1]University of California, Department of Statistics, 367 Evans Hall #3860, Berkeley, CA 94720-3860, USA, e-mail: bradluen@stat.berkeley.edu; stark@stat.berkeley.edu

[2]http://scecdata.usc.edu/csep/

geophysicists disagree about whether earthquake prediction is possible in principle.[3] How, then, ought we decide whether a method for predicting earthquakes works?

Earthquake predictions have been assessed using ideas from statistical hypothesis testing: A test statistic is compared to its distribution under a null hypothesis [10, 11, 19, 22, 28, 40, 55, 57]. The null hypothesis is rejected at significance level $\alpha$ if the test statistic exceeds the $1-\alpha$ quantile of its distribution under the null hypothesis. If the null hypothesis is rejected, researchers tend to conclude — erroneously — that the predictions must have merit.

The null hypothesis can be rejected for many reasons. A Type I error might occur. Or the null hypothesis could be false, but in a way that does not imply that the predictions work. For example, the null hypothesis might use a poor model for seismicity. Or the null hypothesis might not account for how the predictions depend on seismicity. We explore that possibility below.

Conclusions ultimately depend on details of the null hypothesis that can be difficult to justify, or that are known to contradict the data. For example, Freedman and Stark[4] [13] argue that standard interpretations of probability do not make sense for earthquakes–especially for large events, the most important to predict. For rare events, such as large earthquakes, there are not enough data to test or discriminate among competing stochastic models. Models are often calibrated using empirical scaling laws that tie the rates of occurence of large earthquakes to the rates for smaller earthquakes. Generally, these rules of thumb are themselves fitted to data from other parts of the world: applying them to a region as small as the San Francisco Bay area, for example, is questionable. Thus, stochastic models for earthquake occurrence do not seem like a good foundation for evaluating earthquake predictions, especially predictions of large earthquakes.

Moreover, Stark [42, 43] argues that testing predictions using a stochastic model for seismicity and conditioning on the predictions tends to be misleading and that it is preferable to treat seismicity as fixed and compare the success of the predictions with the success of a simple rule. Consider rain forecasts as an analogy. The rule "if it rains today, predict that it will rain tomorrow; otherwise, predict that it will not rain tomorrow" works pretty well. If a meteorologist cannot do better, the meteorologist's predictions have little value.

The seismic analogue is "if there is an earthquake with magnitude greater than the threshold $M_\tau$, predict that there will be an earthquake with magnitude $M$ or above within time $t$ and distance $d$ of the first." Here, $M$, $t$ and $d$ might depend on the location or magnitude of the first earthquake. Kagan [20] uses this "automatic alarm" strategy to evaluate earthquake predictions for Greece (the VAN predictions). The approach can also include a stochastic element to make a "semi-automatic alarm" strategy: Stark [42, 43] compares the VAN predictions to the rule: "If there is an earthquake with magnitude $\geq M_\tau$, toss a (biased) coin. If the coin lands heads, predict that there will be an earthquake with magnitude $\geq M$ within time $t$ and distance $d$ of the first. If the coin lands tails, do not make a prediction."

## 2. Phenomenology of earthquakes

See Bolt [5] for a lay review. The *epicenter* of an earthquake is the point on Earth's surface directly above the earthquake's *focus*, the place that the motion nucleates.

---

[3]See, e.g., http://www.nature.com/nature/debates/earthquake/equake_frameset.html and [14].

[4]A preprint is available at http://statistics.berkeley.edu/tech-reports/611.pdf.

Epicenters and foci are not known exactly: They are estimated from ground motion at seismographic observing stations around the globe. Sizes of earthquakes are also estimated from ground motion measured at seismographic stations. There are many measures of earthquake size, including several definitions of "magnitude."

An earthquake *catalog* is a list of the estimated foci, times, and magnitudes of earthquakes found by a given authority, such as the U.S. Geological Survey. Earthquake catalogs are incomplete below some magnitude (left-censored in magnitude) because smaller events are harder to identify and locate. Moreover, unless some minimum number of stations detect ground motion, the algorithms used to locate earthquakes do not even conclude that there was an earthquake. (The incompleteless as a function of magnitude tends to decrease with time, as equipment becomes more sensitive and networks more extensive.)

Earthquakes occur to depths of 700 km or so in Earth's mantle [45]; however, most earthquakes and almost all large earthquakes occur within a few tens of kilometers of the surface. Earthquakes cluster in space. Most earthquakes occur on pre-existing faults. With very few known exceptions, epicenters of large earthquakes are close to the margins of tectonic plates, because it takes large strains—the relative motions of plates—to produce large earthquakes. Indeed, most large earthquakes occur in a relatively narrow band around the Pacific Ocean, the "ring of fire."

Earthquakes also cluster in time: Large earthquakes almost invariably have aftershocks; some have foreshocks; and there are "swarms" of moderate-to-large earthquakes. Defining foreshocks and aftershocks is difficult. The terms "foreshock" and "aftershock" imply a causal connection to a main shock. Unfortunately, earthquake physics is largely a mystery. Proximity in space and time can be coincidental rather than causal. One cannot tell whether an earthquake is the main shock or a foreshock of a larger event except—at best—in retrospect.[5] And stochastic models for earthquakes can produce spatiotemporal clustering without foreshocks or aftershocks *per se* (for example, Gamma renewal models [48]).

The most common stochastic model for seismicity takes the epicenters and times of shocks[6] above some threshold magnitude to be a realization of a spatially inhomogeneous but temporally homogeneous Poisson process. The spatial heterogeneity reflects tectonics: Some regions are more active seismically than others. The temporal homogeneity is justified by appeal to the lengthy time scale of plate tectonics (tens of thousands of years) relative to the time scale of observation, which is on the order of centuries. The seminal reference on stochastic models for seismicity is Vere-Jones [52], which considers temporal and marked temporal processes, but not spatial processes. Some recent models use branching processes [33, 34, 35].

## 3. Tests of earthquake predictions

There are two categories of earthquake predictions: *deterministic* or *binary predictions*, which are of the form "there will be an earthquake of magnitude $\geq 6$ within 100 km of San Francisco, CA, within the next 30 days;" and *probabilistic predictions*, which are probability distributions, or statements of the form "there is a 90% chance of an earthquake of magnitude $\geq 6$ within 100 km of San Francisco, CA, in the next 30 days." Freedman and Stark [13] point out some difficulties in interpreting probabilistic predictions, using the USGS prediction for the San Francisco Bay Area as an example; here we concentrate on deterministic predictions.

---

[5]Identifying an event as a foreshock or aftershock is a causal inference based on association in time and space. Causal conclusions from associations in non-experimental data are highly suspect. See, e.g., Freedman [12].

[6]Sometimes this is restricted to main shocks, which are difficult to distinguish from foreshocks and aftershocks, as noted above.

To keep the exposition simple, we take the goal to be to predict all earthquakes that exceed some threshold magnitude $M$, that have epicenters in some region $R$ of Earth's surface, and that occur during some time period $T$. We examine several statistical approaches to testing whether predictions have merit.[7]

Let $Q$ denote the total number of earthquakes of magnitude $\geq M$ with epicenters in $R$ during $T$. Let $A$ denote the number of alarms (predictions). The $j$th alarm is characterized by $V_j$, a connected region of space, time and magnitude, and a value $p_j$, the probability the alarm assigns to the occurrence of an earthquake in $V_j$. Deterministic predictions take $p_j = 1$. They assert that an earthquake will occur in $V_j$. Probabilistic forecasts assign a probability $p_j \in (0,1)$ to the occurrence of an earthquake in $V_j$. Let $\delta_j$ be the duration of $V_j$. When $\delta_j$ is on the order of weeks, $V_j$ is generally considered a "prediction." When $\delta_j$ is on the order of a year or more, $V_j$ is generally considered a "forecast." However, some authors use "prediction" to mean deterministic prediction and "forecast" to mean probabilistic prediction, regardless of the time horizon.

Let $\lambda_j$ denote the historical rate of earthquakes in the spatial and magnitude range—but not the temporal range—covered by $V_j$.[8] The historical rates $\{\lambda_j\}_{j=1}^A$ enter into some tests, as we shall see. Let $S_j$ indicate whether the $j$th alarm is successful:

$$
(1) \qquad S_j \equiv \begin{cases} 1, & \text{if there is an earthquake in } V_j, \\ 0, & \text{otherwise}; \end{cases}
$$

and for $k = 1, \ldots, Q$ let $P_k$ indicate whether the $k$th earthquake is predicted:

$$
(2) \qquad P_k = \begin{cases} 1, & \text{if the } k\text{th event is in some } V_j, j = 1, \ldots, A, \\ 0, & \text{otherwise}. \end{cases}
$$

Let $S \equiv \sum_{j=1}^A S_j$ denote the number of successful alarms, and let $P \equiv \sum_{k=1}^Q P_k$ denote the number earthquakes that are predicted successfully. The number of false alarms is $F = A - S$ and the number of earthquakes that are missed—not predicted—is $M = Q - P$. Let $V$ be the volume of space-time studied,

$$
(3) \qquad V \equiv \int_{R,T} dr\,dt,
$$

and let $V_A$ denote the total space-time volume of alarms,

$$
(4) \qquad V_A \equiv \int_{\bigcup_{j=1}^A V_j} dr\,dt.
$$

The fraction of the study volume covered by alarms is $v = V_A/V$. Generally, the smaller $v$ is, the more informative the alarms are, but this can be distorted by spatial heterogeneity of the distribution of earthquakes in $R$.[9] The success rate of the predictions is $s = S/A$; the fraction of earthquakes successfully predicted is $p = P/Q$; the false alarm rate is $f = F/A$; and the rate of missed events (failures to predict) is $m = M/Q$. It we raise an alarm for the entire study volume and time $V$, we can ensure that $s = p = 1$, but then $v = 1$, so the alarms are not informative.

---

[7]Statistical terminology is used in some unfamiliar ways in the geophysical literature. For example, "significance" and "confidence" sometimes denote 100% minus the $P$-value, rather than the chance of a type I error for a fixed-size test (e.g., [28], p. 193 and [9], pp. 723, 731, which also confuses the $P$-value with the chance that the null hypothesis is true). "Random probabilities" are sometimes fixed parameters [19], p. 3773, and "parameters" sometimes means statistics [8], p. 263.

[8]Typically, $\lambda_j$ is the empirical rate over a span of a decade or more over a spatial region that includes $V_j$.

Predictions are generally evaluated using a combination of $s$, $p$, $f$, $m$ and $v$. Prediction methods can be ranked by adjusting their tuning parameters so that their values of $v$ are equal, then comparing their values of $p$, or vice versa. For a given alarm volume $v$, the method with largest $p$ is best. For a given value of $p$, the method with the smallest $v$ is best. Some evaluation strategies fix $p$ and compare values of $f$, or vice versa.

### 3.1. Testing strategies

A common strategy for evaluating earthquake predictions statistically is to compare the success of the predictions on the observed seismicity with the success of the same predictions on random seismicity (e.g., [22, 31, 39, 51]). This strategy does not make sense because predictions usually depend on past seismicity: If the seismicity had been different, the predictions would have been different.[10]

Several stochastic models for seismicity are common in null hypotheses for testing predictions.

1. Some studies model seismicity by a homogeneous Poisson process with intensity equal to the mean historical rate in the study region (e.g., [10]). Some studies condition on the number of events and model seismicity as uniform over the study region or subsets of the study region [6, 32, 55].
2. Some studies use a spatially heterogeneous but temporally homogeneous Poisson process model, with rate in the spatial region $R_j$ equal to the historical rate $\lambda_j$ [11, 28].
3. Some studies condition on the observed locations of past events, but model the times of the events as Poisson or uniform [28, 47].
4. Some studies condition on the observed locations and the observed times, but model the times as exchangeable [20, 37]. That is, if the observed time of the $j$th event in the catalog is $t_j$, $j = 1, \ldots, Q$, then, according to the model, it is equally likely that the times would have been $t_{\pi(j)}$, $j = 1, \ldots, Q$, where $\pi$ is any permutation of $\{1, \ldots, Q\}$.

In the last approach (the permutation model, sometimes called "randomizing a catalog"), times of events in the study region are exchangeable.

There are variations on these approaches. For example, some researchers try to remove putative aftershocks from the catalogs (e.g., [1, 20, 40]). This is called

---

[9]To account for the spatial heterogeneity of events, some authors use normalized counting measure in space — based on the historical occurrence of events in a given volume — rather than Lebesgue measure. See, e.g., Kossobokov et al. [28].

[10]This is a bit like the Monte Hall or *Let's Make a Deal* problem [44], Chapter 10: A prize is hidden at random behind one of three doors. The contestant picks a door. The host then reveals that the prize is not behind one of the two doors the contestant did not pick. The contestant is now allowed to switch his guess to the third door. Should he? Some erroneous arguments assume that the door the host opens is independent of which door conceals the prize. That is not a good model for the game because the host never opens the door that hides the prize: Which door the host opens depends on the contestant's guess and on which door hides the prize. Similarly, holding the prediction fixed regardless of the seismicity is not a good model for earthquake prediction. Whether a prediction is issued for tomorrow typically depends on whether there is an earthquake today.

"declustering." The most common method for declustering is to make spatiotemporal holes in a catalog: After each event, all smaller events that occur within a given time interval and epicentral distance are deleted. The time interval and distance can depend on the magnitude of the event [23, 27, 38]. (For an alternative approach to declustering, see [56].) It is common to assume that a declustered catalog is a realization of a temporally homogeneous Poisson process.[11] Assessments of earthquake predictions are known to be sensitive to details of declustering and to spatial variability of the rate of seismicity [17, 21, 29, 31, 42, 43, 49, 54].

Another approach to testing is to compare the success rate of predictions with the (theoretical or empirical) success rate of random predictions that do not use any seismic information [16]. This seems to be a straw-man comparison because such random predictions ignore the empirical clustering of seismicity.

### 3.2. Jackson, 1996

Jackson [19] reviews methods for testing deterministic and probabilistic predictions. The approach to testing deterministic predictions is based on a probability distribution for the number of successful predictions, in turn derived from a null hypothesis that specifies $\mathbb{P}(S_j = 1)$, $j = 1, \ldots, A$. Jackson does not say how to find these probabilities, although he does say that usually the null hypothesis is that seismicity follows a Poisson process with rates equal to the historical rates. He assumes that $\{S_j\}_{j=1}^A$ are independent, so $S$ is the sum of $A$ independent Bernoulli random variables. Jackson advocates estimating the $P$-value, $\mathbb{P}(S \geq S_{\text{observed}})$, by simulating the distribution of the sum of independent Bernoulli variables, and mentions the Poisson approximation as an alternative. See Kagan and Jackson [22] for more discussion of the same approaches. Both articles advocate a likelihood-ratio test for evaluating probabilistic forecasts. (See [16] for an application.) They also propose a variant of the Neyman-Pearson testing paradigm in which it is possible that both the null hypothesis and the alternative hypothesis are rejected, in effect combining a goodness-of-fit test of the null with a likelihood ratio test against the alternative.

### 3.3. Console, 2001

Console [8] addresses deterministic predictions and probabilistic forecasts. His discussion of deterministic predictions includes several statistics for comparing alternative sets of predictions. His discussion of probabilistic forecasts is based on the likelihood approach in Kagan and Jackson [22], described above. The likelihood function assumes that predictions succeed independently, with known probabilities. For Console, the null hypothesis is that seismicity has a Poisson distribution ([8], page 266). He gives one numerical example of testing a set of four predictions on the basis of "probability gain," but no hint as to how to determine the significance level or power of such tests. His test rejects the null hypothesis if more events occur during alarms than are expected on the assumption that seismicity has a homogeneous Poisson distribution with true rate equal to the observed rate. Console also mentions selecting prediction methods on the basis of a risk function, and Bayesian methods. The loss function Console contemplates is linear in the number of predicted events, the number of unpredicted events, and the total length of alarms, all of which are treated as random. He does not address estimating the risk from data, but it seems that any estimate must involve stochastic assumptions about $Q$, $S$, $F$ and $M$.

---

[11]This kind of declustering produces a process that has less clustering than a Poisson process because it imposes a minimum distance between events.

### 3.4. Shi, Liu and Zhang, 2001

Shi, Liu and Zhang [40] evaluate official Chinese earthquake predictions of earthquakes with magnitude 5 and above for 1990–1998. They divide the study region into 3,743 small cells in space, and years of time. In a given cell in a given year, either an earthquake is predicted to occur, or — if not — that is considered to be a prediction that there will be no event in that cell during that year. They define the $R$-score as

$$R = \frac{\#\text{ cells in which earthquakes are successfully predicted}}{\#\text{ cells in which earthquakes occur}}$$
$$(5) \qquad - \frac{\#\text{ cells with false alarms}}{\#\text{ aseismic cells}},$$

which measures the concordance of the binned data with predictions of occurrence and of non-occurrence. In computing the $R$-score, they first decluster the catalog using the method of Keilis-Borok et al. [23]. Their hypothesis tests use the $R$-score as the test statistic. They compare the $R$-score of the actual predictions on the declustered catalog with the $R$-score of several sets of random predictions, generated as follows:

1. Condition on the number of cells in which earthquakes are predicted to occur. Choose that many cells at random without replacement from the 3,743 cells, with the same chance of selecting each cell; predict that earthquakes of magnitude 5 or above will occur in those randomly-selected cells.
2. To take spatial heterogeneity into account, for the $j$th cell, toss a $p_j$-coin, where $p_j$ is proportional to the historical rate of seismicity in that cell. If the $j$th coin lands heads, predict that an earthquake of magnitude 5 or above will occur in the $j$th cell. Toss coins independently for all cells, $j = 1, \ldots, 3743$. The constant of proportionality is the ratio of the number of cells for which the actual predictions anticipates events, divided by the historical annual average number of cells in which events occur. This produces a random number of predictions, with predictions more likely in cells where more events occurred in the past.
3. Condition on the number of cells in which earthquakes are predicted to occur. Choose that many cells at random without replacement from the 3,743 cells. Instead of selecting cells with equal probability, select the $j$th cell with probability $p_j$, with $p_j$ set as in (2). Predict that earthquakes of magnitude 5 or above will occur in those randomly-selected cells.

The third approach is a blend of the first two approaches: The number of simulated predictions each year is forced to equal the actual number of predictions, but the chance of raising a prediction in the $j$th cell depends on the historical rate of seismicity in the $j$th cell. None of these three comparison methods depends on the observed seismicity during the study period, 1990–1998. In particular, none exploits clustering, which is presumed to have been eliminated from the catalog.

## 4. Some claims of successful predictions

### 4.1. Wyss and Burford, 1987

Wyss and Burford [55] claim to have predicted the magnitude $M_L = 4.6$ earthquake that occurred on 31 May 1986 near Stone Canyon, California, about a year before it occurred, using "seismic quiescence," an anomalous paucity of earthquakes over some period of time. They examine the rates of earthquakes on different sections of the San Andreas fault and identify two fault sections in which the rate dropped compared with the rates in neighboring sections. They say that "the probability [of the prediction] to have come true by chance is $< 5\%$." The probability they calculate is the chance that an earthquake would occur in the alarm region, if earthquakes occurred at random, independently, uniformly in space and time, with rate equal to the historic rate in the study area over the previous decade. In effect, their null hypothesis is that seismicity follows a homogeneous Poisson process with rate equal to the historical rate; clustering is not taken into account.

### 4.2. VAN predictions based on Seismic Electrical Signals

There has been a lively debate in the literature about whether predictions made by Varotsos, Alexopoulos and Nomicos (VAN) [50] of earthquakes in Greece succeeded beyond chance. See volume 23 of *Geophysical Research Letters* (1996). The participants did not even agree about the number of earthquakes that were predicted successfully, much less whether the number of successes was surprising. Participants disagreed about whether the predictions were too vague to be considered predictions, whether some aspects of the predictions were adjusted *post hoc*, what the null hypothesis should be, and what tests were appropriate.

### 4.3. Kossobokov et al., 1999

Kossobokov, Romashkova, Keilis-Borok and Healy [28] claim to have predicted four of the five magnitude 8 and larger earthquakes that occurred in the circum-Pacific region between 1992 and 1997. They say "[t]he statistical significance of the achieved results is beyond 99%." (From context, it is clear that they mean that the $P$-value is $< 1\%$.) Their predictions are based on two algorithms, M8 and MSc, which track the running mean of the number of main shocks; the difference between the cumulative number of main shocks and a windowed trend in the number of main shocks; a measure of spatial clustering of main shocks derived from the distance between shocks and the diameters of the sources in a temporal window; and the largest number of aftershocks of any event in a temporal window. These are used to identify "times of increased probability," which are predictions that last five years. The declustering method described above was used to classify events as main shocks or aftershocks.

Kossobokov et al. [28] calculate statistical significance by assuming that earthquakes follow a Poisson process that is homogeneous in time but heterogeneous in space, with an intensity estimated from the historical rates of seismicity, $\{\lambda_j\}$. Kossobokov et al. [28] condition on the number of events that occur in the study area, which leads to a calculation in which locations and times are iid across events, the epicenters and times are independent of each other, the temporal density of earthquake times is uniform, and the spatial distribution of epicenters is given by the historical distribution between 1992 and 1997. Their calculation does not take temporal clustering into account, and it conditions on the predictions. They calculate the chance that $S$ or more of the $Q$ events would occur during alarms to be $\sum_{x=S}^{Q} {}_Q C_x \pi^x (1-\pi)^{S-x}$, where $\pi$ is the normalized measure of the union of the alarms. The measure is the product of the uniform measure on time and counting measure on space, using the historical distribution of epicenters in the study volume to define the counting measure.

## 5. A naive predictor

In this section we exploit the empirical clustering of earthquakes to construct a predictor that succeeds far beyond chance according to some tests that hold predictions fixed and treat seismicity as random. The chance model for seismicity uses the observed times and locations of earthquakes, but shuffles the times. That is, according to the null hypothesis, the times of events are exchangeable given their locations and magnitudes and the predictions. We can simulate from the null model by randomly permuting the list of observed times relative to the list of observed locations and magnitudes. This is a common "nonparametric" approach to testing earthquake predictions [20, 37]. That is, if the location, magnitude and time of the events in the catalog are $\{(r_j, M_j, t_j)\}_{j=1}^{Q}$, we take the $Q!$ outcomes $\{(r_j, M_j, t_{\pi(j)})\}_{j=1}^{Q}$ (as $\pi$ ranges over all $Q!$ permutations of $\{1, \ldots, Q\}$) to be equally likely under the null hypothesis, given the predictions. We do not claim that this predictor or this null hypothesis is good. Rather, we claim that this approach to testing is misleading.

We apply the approach to the Harvard centroid moment tensor (CMT) catalog[12] for 2004 and for 2000–2004. We make two sets of predictions:

(i) After each earthquake of (body-wave) magnitude $M_\tau$ or greater, predict that there will be an earthquake of magnitude $M_\tau$ or greater within 21 days, and within 50 km epicentral distance.
(ii) After each earthquake of magnitude $M_\tau$ or greater, predict that there will be an earthquake within 21 days and within 50 km that is at least as large as any within 50 km within 21 days prior to its occurrence.

Predictor (ii) is equivalent to predictor (i) if an event is deemed eligible for prediction only if there is no larger event within 50 km in the 21 days leading up to the event.

Let $M_j$ be the magnitude of the $j$th event that triggers an alarm; let $t_j$ be the time of the $j$th event that triggers an alarm; and let $R_j$ be the set of points on Earth's surface that are within 50 km of the epicenter of the $j$th event that triggers an alarm. Recall that an alarm is a connected region $V_j$ of space, time, and magnitude. For predictor (i),

(6) $$V_j = R_j \times (t_j, t_j + 21 \text{ days}] \times [M_\tau, \infty),$$

while for predictor (ii),
(7)
$$V_j = \{R_j \times (t_j, t_j + 21 \text{ days}] \times [M_j, \infty)\} \setminus \bigcup_{k: M_k > M_j} \{R_k \times (t_k, t_k + 21 \text{ days}] \times [M_j, M_k)\}.$$

---

[12]http://www.globalcmt.org/CMTsearch.html.

*Testing earthquake predictions* 311An event at time $t$, epicenter $r$ and with magnitude $M$ is predicted by the second set of predictions if and only if

(8) $$M \in \bigcap_j \{[M_j, \infty) : (t, r) \in (t_j, t_j + 21 \text{ days}] \times R_j\}.$$

Predictor (i) tends to predict more aftershocks: Large events trigger alarms that contain some of their aftershocks. Predictor (ii) prevents aftershocks from being predicted by main shocks; however, it does not prevent aftershocks with magnitude $M_\tau$ or larger from predicting still larger aftershocks of the same main event, provided the predicted aftershock is the largest event in the preceding 21 days, within 50 km. Predictors (i) and (ii) generate the same number of alarms and have the same total duration, but not the same extent of space and magnitude. Note that these alarms need not be disjoint.

We consider two values of the threshold magnitude $M_\tau$: 5.5 and 5.8. We compare the number of events successfully predicted by these two predictors with the distribution of number that would be predicted successfully if seismicity were "random." Using the CMT catalog, we generate a set of alarms. Holding those alarms fixed, we see how successful the alarms would be in predicting random seismicity — generated by randomly permuting the times in the CMT catalog.

Table 1 summarizes the results. Under the null hypothesis, both prediction methods (i and ii) succeed well beyond chance for CMT data from the year 2004 and the years 2000–2004, for both values of the threshold magnitude. That is not because the prediction method is good; rather, it is because the stochastic model in the null hypothesis fails to take into account the empirical clustering of earthquakes and the dependence of the predictions on the seismicity. Holding predictions fixed as seismicity varies randomly does not make sense.

TABLE 1
*Simulation results using the global Harvard Centroid Moment Tensor (CMT) catalog. We seek to predict events with body-wave magnitude $M_\tau$ and above. "Events" is the total number of events in the time period with magnitude at least $M_\tau$. Each event with body-wave magnitude $M_\tau$ or greater triggers an alarm. In each row, the number of alarms is equal to the number of events in column 3. The spatial extent of the alarm is a spherical cap of radius 50 km centered at the epicenter of the event that triggers the alarm. The temporal extent of the alarm is 21 days, starting at the time of the event that triggers the alarm. We set the magnitude extent of alarms in two ways. Column 4, 'succ,' is the number of successful predictions using predictor (i): It is the number of events with magnitude at least $M_\tau$ that are within 21 days following and within 50 km of the epicenter of an event with magnitude $M_\tau$ or greater. Column 5, 'succ w/o,' is the number of successful predictions using predictor (ii): It is the number of events that are within 21 days following and within 50 km of the epicenter of an event whose magnitude is at least $M_\tau$ but no greater than that of the event in question. Events that follow within 21 days of a larger event are not counted. Column 6, 'max sim,' is the largest number of successful predictions in 1,000 random permutations of the times of the events Harvard CMT catalog, holding the alarms and the locations and magnitudes of events in the catalog fixed. The alarms are those corresponding to column 5 — predictor (ii) in the text — that is, an event is eligible for prediction only if its magnitude exceeds that of every event within 50 km within the 21 days preceding it. Column 7, 'P-value (est),' is the estimated P-value: the fraction of permutations in which the number of successful predictions was greater than or equal to the observed number of successful predictions for the CMT catalog. This corresponds to predictor (ii), which is intended to reduce the number of predictions satisfied by aftershocks. Column 8, 'v,' is an upper bound on the fraction of the study region (in space and time) covered by alarms; it is not adjusted for overlap of alarms.*

| year | $M_\tau$ | events | succ | succ w/o | max sim | $P$-value (est) | $v$ |
|---|---|---|---|---|---|---|---|
| 2004 | 5.5 | 445 | 95 | 30 | 28 | <0.001 | $3.9 \times 10^{-4}$ |
| 2004 | 5.8 | 207 | 24 | 7 | 10 | 0.041 | $1.8 \times 10^{-4}$ |
| 2000–2004 | 5.5 | 2013 | 320 | 85 | 48 | <0.001 | $3.6 \times 10^{-4}$ |
| 2000–2004 | 5.8 | 996 | 114 | 29 | 19 | <0.001 | $1.8 \times 10^{-4}$ |

## 6. Conclusions

Interpreting earthquake predictions is difficult. So is evaluating whether predictions work. To use a statistical hypothesis test, something must be assumed to be random, and its probability distribution under the null hypothesis must be known. Many studies model seismicity as random under the null hypothesis. That approach has serious drawbacks, and details of the stochastic model, such as spatial heterogeneity, independence or exchangeability, matter for testing. Most null hypotheses used in tests ignore the empirical clustering of earthquakes. Some try to remove clustering with ad hoc adjustments as a prelude to probability calculations. It is implausible that the resulting data represent a realization of a Poisson process, as is often assumed. The standard approach to testing — hold the predictions fixed while seismicity varies randomly according to some stochastic model — does not take into account that in practice, the predictions would be different if the seismicity were different. The result is that simple-minded schemes, such as the "automatic alarm strategy," can succeed well beyond chance in hypothesis tests. This is not because the predictions are good: It is because the tests are bad.

**Acknowledgments.** We are grateful to David A. Freedman, Cliff Frohlich, Robert Geller and Yan Y. Kagan for helpful conversations.